\documentclass[a4paper,fleqn,usenatbib]{mnras}

\usepackage[T1]{fontenc}
\usepackage{ae,aecompl}

\usepackage{graphicx}	
\usepackage{amsmath}	
\usepackage{amssymb}	
\usepackage{pdflscape}

\newcommand{\kms} {$\mathrm{ km \; s^{-1}}\,$}
\newcommand{\msol} {M$_{\odot}$}
\newcommand{\mza} {M$_{ZAMS}$}

\newcommand{\about} {$\sim$}

\def\lesssim{\mathrel{\hbox{\rlap{\hbox{\lower4pt\hbox{$\sim$}}}\hbox{$<$}}}}
\def\gtrsim{\mathrel{\hbox{\rlap{\hbox{\lower4pt\hbox{$\sim$}}}\hbox{$>$}}}}

\newcommand{\degree}{$^{\circ}$}

\title[Spectropolarimetry of SN 2011hs]{The 3D shape of Type IIb SN 2011hs}

\author[H. F. Stevance et al.]{H. F. Stevance$^{1}$\thanks{E-mail: fstevance1@sheffield.ac.uk},  J.R. ~Maund$^{1}$\thanks{Royal Society Research Fellow},  D. ~Baade$^{2}$, J. Bruten$^{1}$,  A. Cikota$^{3}$, P. ~H\"oflich$^{4}$, \newauthor L. ~Wang$^{5}$, J.C. ~Wheeler$^{6}$,     A. Clocchiatti$^{7}$, J. ~Spyromilio$^{2}$,  F. ~Patat$^{2}$,  Y. ~Yang$^{8}$, \newauthor P. Crowther$^{1}$
\\
$^{1}$University of Sheffield, Department of Physics and Astronomy, Hounsfield Rd, Sheffield S3 7RH, UK., \\
$^{2}$European Organisation for Astronomical Research in the Southern Hemisphere (ESO), Karl-Schwarzschild-Str. \\ \,\,\, 2, D-85748 Garching b. M{\"u}nchen, Germany,\\  
$^{3}$Physics Division, Lawrence Berkeley National Laboratory, 1 Cyclotron Road, Berkeley, CA 94720, USA\\
$^{4}$Department of Physics, Florida State University, Tallahassee, FL 32306-4350, USA,\\
$^{5}$George P. and Cynthia Woods Mitchell Institute for Fundamental Physics \& Astronomy, Texas A. \& M. University,\\
$^{6}$Department of Astronomy and McDonald Observatory, The University of Texas at Austin, Austin, TX 78712, USA,\\
$^{7}$Institute of Astrophysics, Universidad Cat\'{o}lica de Chile, and Millennium Institute of Astrophysics, Santiago, Chile\\
$^{8}$2 Department of Particle Physics and Astrophysics, Weizmann Institute of Science, Rehovot 76100, Israel.\\
}

\date{Accepted XXX. Received YYY; in original form ZZZ}

\pubyear{2018}

\begin{document}
\label{firstpage}
\pagerange{\pageref{firstpage}--\pageref{lastpage}}
\maketitle

\begin{abstract}
We observed seven epochs of spectropolarimetry in optical wavelengths for the Type IIb SN 2011hs, ranging from $-$3 to +40 days with respect to $V$-band maximum. 
A high degree of interstellar polarization was detected (up to \about3 percent), with a peak lying blueward of 4500\r{A}. 
Similar behaviours have been seen in some Type Ia SNe, but had never been observed in a Type IIb. 
We find that it is most likely the result of a relative enhancement of small silicate grains in the vicinity of the SN.
Significant intrinsic continuum polarization was recovered at $-$3 and +2 days (p = 0.55 $\pm$ 0.12 percent and p = 0.75 $\pm$ 0.11 percent, respectively). 
We discuss the change of the polarization angle across spectral lines and in the continuum as diagnostics for the 3D structure of the ejecta.
We see a gradual rotation by about $-$50\degree\, in the continuum polarization angle between $-2$ and +18 days after $V$-band maximum. 
A similar rotation in He\,{\sc i} $\lambda 5876$, $\mathrm{H\alpha}$ and the Ca\,{\sc ii} infrared triplet seems to indicate a strong influence of the global geometry on the line polarization features. 
The differences in the evolution of their respective loops on the Stokes $q-u$ plane suggest that line specific geometries are also being probed.
Possible interpretations are discussed and placed in the context of literature.
We find that the spectropolarimetry of SN 2011hs is most similar to that of SN 2011dh, albeit with notable differences.
\end{abstract}




\section{Introduction}
\label{intro}
At the end of their lives, the cores of massive stars (M$_{\rm ZAMS} >$ 8 M$_{\odot}$) collapse, which can result in core collapse supernovae (CCSNe).
There are multiple types of CCSNe: Type II SNe show strong hydrogen features in their spectra, whereas Type Ib/c SNe do not.
Type Ib and Ic SNe are differentiated based on the presence (Ib) or absence (Ic) of helium in their spectra. 
There also exists a transitional type of SN that bridges the gap between Type II and Type Ib SNe: the Type IIb, whose spectrum evolves from being hydrogen dominated (Type II) to being helium dominated (Ib -- \citealt{filippenko97, branch17}).
Type IIb SNe arise from progenitors that have been stripped of nearly all of their hydrogen envelope, retaining less than 0.5 M$_{\odot}$ \citep{smith11}. 
This stripping can either be caused by mass transfer in a binary system or result from mass-loss through strong stellar winds.
As such, Type IIb SNe are sensitive probes of mass loss processes, particularly of binary interactions (e.g \citealt{maund93J}, \citealt{fox14}).

A number of progenitors of CCSNe have been imaged (see \citealt{smartt09} for a review) and explosion models are extensively discussed in the literature (e.g see the reviews of \citealt{janka12} and \citealt{burrows13}). 
SNe in other galaxies are unresolved at early times, but spectropolarimetry is a unique tool that can probe the 3D shape of young SN ejecta and thus provide constraints on the explosion models. 

The opacity of the photosphere at early days is dominated by electron scattering \citep{ss82}. 
As a result, the outgoing radiation has a polarization vector perpendicular to the plane of last scattering (i.e containing the incident and scattered ray). 
Additionally, the amplitude of the polarization quasi-vector increases with distance to the centre from the ejecta projected onto the plane of the sky \citep{chandra46}.
In a spatially unresolved envelope with central symmetry (e.g spherical), the polarization components will cancel fully, resulting in null integrated polarization.
Departure from spherical symmetry, however, will result in incomplete cancellation and a polarization excess \citep{ss82,mccall84}, which can be measured to probe the shape of the SN ejecta. 
Deviations from central symmetry can be the result of a photospheric geometry \citep{hoflich91}, off-centre energy sources \citep{chugai92, hoflich95}, or partial occultation of the underlying photosphere by a non-isotropic distribution of the line forming regions (e.g. \citealt{kasen03}). 
Additionally, non-zero polarization can be the result of scattering by circumstellar dust \citep{wang96}.

Practically all CCSNe show net intrinsic polarization \citep{WW08}.
Despite their relatively low rates (\about 12 percent of CCSNe, \citealt{eldridge13}), Type IIb SNe are surprisingly well represented in the spectropolarimetry literature and have shown a range of behaviours.
Most Type IIb SNe -- such as SN 1993J, SN 1996cb, SN 2008ax, SN 2008aq, or SN 2011dh -- have continuum polarization around $p \sim$ 0.5 percent to $p \sim$ 1 percent \citep{trammell93,wang01,chornock11,silverman09,stevance16,mauerhan15}.
Some Type IIb SNe (e.g. SN 2001ig or SN 2008aq) exhibited a spectropolarimetric evolution similar to that of Type II-P SNe, where a lower continuum polarization was seen at early times and increased to $p >1$ percent by the time helium started dominating the spectrum \citep{ leonard06, maund01ig, stevance16}. 
SN 2011dh, on the other hand, exhibited a decrease in polarization from \about0.75 percent around maximum light to \about 0.2 percent by 8 days after maximum.
Additionally, strong line polarization of hydrogen, helium and calcium features is often detected in Type IIb SNe.
The amplitude of the peaks can range from \about 0.5 percent (e.g. SN 2001ig or SN 2011dh at early times) to above 3 percent in SN 2008ax \citep{maund01ig,mauerhan15,chornock11}.

Although Type IIb SNe are better represented than other types in the spectropolarimetric literature, the number of high-quality multi-epoch data sets remains limited.
Increasing our sample is ultimately necessary to compare the statistical distribution of the observed polarization to available models (e.g. \citealt{hoflich91, kasen03, dessart11, tanaka17}).

In this work we present seven epochs of spectropolarimetric and spectroscopic data for SN 2011hs observed with the Very Large Telescope (VLT) from $-$3 days to +40 days with respect to $V$-band maximum (22 November 2011 -- see \citealt{bufano14}).
The observations are described in Section \ref{sec:obs}.
A study of the interstellar polarization (ISP) in the line of sight of SN 2011hs is given in Section \ref{sec:isp}; we identify unusual levels of ISP and discuss potential implications for the dust in the host galaxy or in the vicinity of the SN. 
In Section \ref{sec:pol} we analyse the intrinsic polarization of SN 2011hs.
In Section \ref{sec:disc}, we compare the intrinsic polarization of SN 2011hs to that of previously studied Type IIb SNe and discuss potential interpretations. 
Finally, a summary is given in Section \ref{sec:conclusions}. 

\section{Observations and data reduction}
\label{sec:obs}
\begin{table}
\centering
\caption{\label{tab:obs} VLT Observations of SN~2011hs.  The epochs are given relative to the estimated V-band maximum.  $^a$Flux Standard. }
\begin{tabular}{c c c c c}
\hline
Object & Date & Exp. Time & Epoch & Airmass \\
 & (UT) & (s) & (days) & \\
\hline

SN~2011hs & 2011 Nov. 19 & 8 $\times$ 850 & $-3$ & $1.05-1.23$ \\
EGGR 141$^a$  & 2011 Nov. 19 & 2$\times$60 & $-$ & 1.26\\
 \\ 
SN~2011hs & 2011 Nov. 24 & 8 $\times$ 900 & +2 &  $1.1-1.36$ \\
EGGR 150$^a$& 2011 Nov. 24 & 2 $\times$ 191 & $-$ & 1.31 \\
 \\
SN~2011hs & 2011 Dec. 02 & 8 $\times$ 900 & +10 & $1.16-1.56$ \\
LTT1788$^a$ & 2011 Dec. 02 & 2 $\times$ 60 & $-$ & 1.04 \\
 \\
SN~2011hs & 2011 Dec. 10 & 8 $\times$ 900 & +18 & $1.3-2.0$ \\
LTT1788$^a$  & 2011 Dec. 10  & 2 $\times$ 60  & $-$ & 1.8 \\
 \\
SN~2011hs & 2011 Dec. 16 & 4 $\times$ 855 & +24 & $1.4-2.3$ \\
LTT1788$^a$ & 2011 Dec. 16 & 2 $\times$ 120 & $-$ & 1.14 \\
  \\
SN~2011hs & 2011 Dec. 23  & 4 $\times$ 1100 & +31 & $1.4-1.77$ \\
LTT1788$^a$ & 2011 Dec. 23 & 2 $\times$ 60  & $-$ & 1.034 \\
 \\
SN~2011hs & 2012 Jan. 01  & 4 $\times$ 1100  & +40 & $1.6-2.1$ \\
LTT1788$^a$  & 2011 Dec. 16 & 2 $\times$ 120 & $-$ & 1.14 \\
\hline
\end{tabular}
\end{table}

SN 2011hs was discovered by Stuart Parker on 12.476 November 2011 and classified by \cite{2011hs}.
It is located at R.A. = 22$^{\text{h}}$57$^{\text{m}}$11$^{\text{s}}$ and $\delta$ = -43\degree23'04" in the galaxy IC 5267 (face-on sA0) with redshift z = 0.005711, corresponding to a recessional velocity of 1714 \kms \citep{koribalski04}.
A series of spectropolarimetric observations of SN 2011hs were taken with the VLT of the European Southern Observatory (ESO) using the Focal Reducer and low-dispersion Spectrograph (FORS2) in the dual-beam spectropolarimeter mode (PMOS -- \citealt{appenzeller98}). 
Linear spectropolarimetry was obtained for 4 half-wave retarder angles  (0\degree, 22.5\degree, 45\degree, 67.5\degree) at seven epochs between 19 November 2011 and 01 January 2012. 
A summary of observations is given in Table \ref{tab:obs}.
All observations were performed with the 300V grism, providing a spectral resolution of 12 \r{A} (as measured from arc-lamp calibration images). 
The GG435 order sorting filter was used to avoid contamination of our data by higher dispersion orders.
Our data are thus limited to the wavelength range 4450\r{A} -- 9330\r{A}
The spectropolarimetric data were reduced in the standard manner using IRAF\footnote[2]{IRAF is distributed by the National Optical Astronomy Observatory, which is operated by the Association of Universities for Research in Astronomy (AURA) under a cooperative agreement with the National Science Foundation.} following the prescription of \cite{maund05bf} and the Stokes parameters were calculated using FUSS \citep{stevance17}. 
The data acquired on 19 November, 24 November and 02 December 2011 were binned to 15 \r{A}, the data obatined on the 10 December 2011 were binned to 30\r{A} and the data obtained at subsequent epochs were binned to 45 \r{A} to increase the signal-to-noise ratio. 
The spectra were flux calibrated using the standard stars reported in Table \ref{tab:obs} with the polarimetric optics in place in order to remove the instrumental response.
It should be noted, however, that due to unknown slit losses, absolute flux calibration was not possible.

\subsection{Observed flux and spectropolarimetry}
\subsubsection{Flux spectrum and photospheric velocity}
\begin{figure}
	\includegraphics[width=8cm]{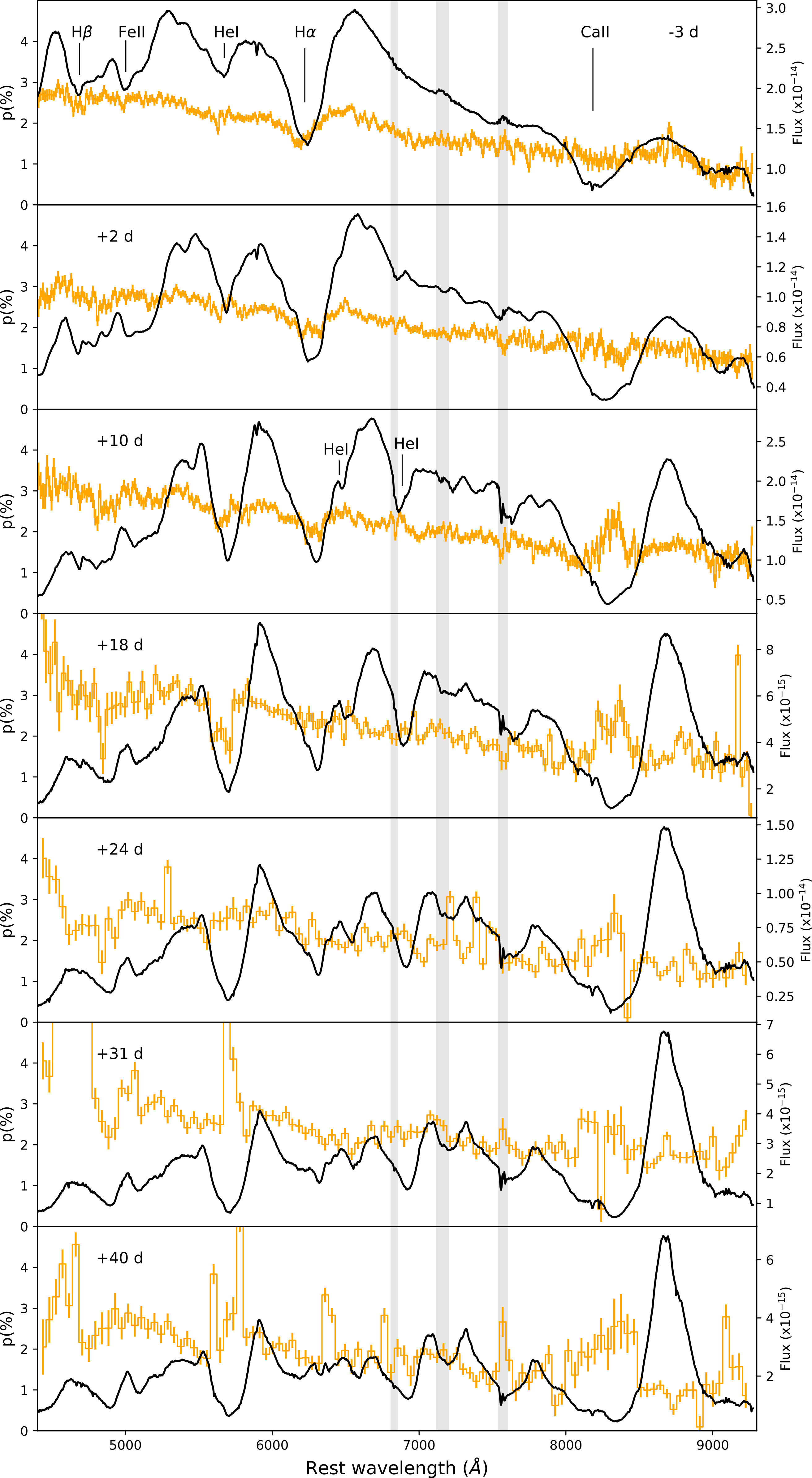}
    \caption{\label{fig:flu_n_pol0}Flux spectrum (black)  and degree of polarization (not corrected for ISP - orange) of SN 2011hs  ranging from $-$3 days to +40 days. The grey shaded areas represent the regions affected by telluric lines. The data at $-$3, +2 and +10 days were binned to 15\r{A}, the data at +18 days were binned to 30\r{A} and the following epochs were binned to 45\r{A}. Note that the strong peaks of polarization near 4600\r{A} and 5700\r{A} at +31 and +40 days are not considered to be real (see Section \ref{sec:obs_pol}). }
\end{figure}

In Figure \ref{fig:flu_n_pol0}, we present the observed spectropolarimetry and flux spectrum at seven new epochs for SN 2011hs.

As noted by \cite{bufano14}, early epochs of spectroscopy are dominated by $\mathrm{H\alpha}$ ($-3$ days), but as the supernova evolves through maximum light the feature decreases in strength and  a blue shoulder appears (+2 days).
We also see that from a week after maximum, the spectrum is dominated by He\,{\sc i} features. 
An exhaustive spectroscopic analysis is beyond the scope of this paper, and more details on the spectroscopy of SN 2011hs and comparison to other SNe can be found in \cite{bufano14}.
Nonetheless, for the needs of this work we estimate the photospheric velocities at the time of observations (see Section \ref{sec:pol}).
The velocities of the Fe\,{\sc ii} lines in the blue part of the spectrum can be used as a proxy to probe photospheric velocity (e.g \citealt{bufano14}), so we perform fits of Fe\,{\sc ii} $\lambda5169$ at all epochs in order to find the absorption minima. 
The resulting velocities are given in Table \ref{tab:vel_table}, and are consistent with the estimates of \cite{bufano14} from fits of the same line, see their figure 9.

\begin{table}
\begin{tabular}{r c }
\hline 
Epoch  & Fe\,{\sc ii}  \\
\hline 
$-$3 days & $-$9 860 \kms \\ 
+2 days & $-$8 760 \kms \\ 
+10 days & $-$6 320 \kms \\ 
+18 days & $-$5 570 \kms \\ 
+24 days & $-$5 390 \kms  \\  
+31 days & $-$4 810 \kms \\ 
+40 days & $-$4 470 \kms\\ 
\hline
\end{tabular}
\caption{\label{tab:vel_table}Line velocities at the absorption minimum for Fe\,{\sc ii} $\lambda 5169$ at all epochs,  which is used as a proxy for photospheric velocity.}
\end{table}

\subsubsection{Observed polarization}
\label{sec:obs_pol}

As seen in Figure \ref{fig:flu_n_pol0}, we detect significant polarization in the direction of SN 2011hs, showing an underlying slope with a higher degree of polarization at the blue end of the spectrum (\about3 percent) than at  the red end (\about1 percent).
This behaviour is not expected for the continuum polarization of SNe since Thomson scattering is wavelength independent.
Additionally, some excess polarization is associated with the absorption component of Ca\,{\sc ii} from 10 days after V-band maximum, and polarization troughs are seen corresponding to $\mathrm{H\alpha}$ at the first 3 epochs, as well as with He $\lambda5876$ at +10 and +18 days.
The presence of troughs is inconsistent with the expected correlation of polarization peaks with absorption components (e.g \citealt{WW08}).
These discrepancies seem to indicate that interstellar polarization (ISP) contributes very significantly to the continuum polarization observed (see Section \ref{sec:isp} for a more complete discussion). 

Lastly, it should be noted that the strong peaks near 4600\r{A} and 5700\r{A} at +31 and +40 days are coincident with significant variations from zero in our instrumental signature correction $\Delta \epsilon$ (see \citealt{maund08}), reducing our confidence in our data at these wavelengths. 
Spectral extraction using the non-optimal method was attempted to test whether a failure of the optimal extraction algorithm could be responsible for such wide anomalies. 
This method, however, showed the same peaks. 
A similar spurious feature over a wide wavelength range has also been encountered in the spectropolarimetric data of WR102 \citep{stevance18}, but its origin could not be identified. 
We do not consider these peaks to be intrinsic to the SN or to the interstellar polarization, and they will therefore not be considered in our analysis.


\section{Interstellar polarization}
\label{sec:isp}

\subsection{Galactic ISP}
\label{sec:gal_isp}
In order to estimate the contribution of the Galactic ISP we searched for Milky Way field stars in the vicinity of SN 2011hs. 
If we assume that they are intrinsically unpolarized, any polarization measured for these stars is due to Galactic ISP. In the \cite{Heiles} catalogue, we found two stars within 2 degrees of SN 2011hs: HD218227 and HD215544 with polarization $p = 0.012 $ and $p = 0.116$ percent, respectively. 
Using the parallaxes found in the Gaia data release DR2 \citep{gaia_parallaxes} we calculate distances of 36.1$\pm0.7$ pc and 501$\pm10$ pc for HD218227 and HD215544, respectively.
HD218227 being such a nearby object explains its low polarization degree compared to HD215544, and the latter better samples the Galactic dust column and the ISP in the direction of SN 2011hs.

It is possible to put an upper limit on the Galactic ISP following $p_{\text{max}} < 9 \times E(B-V)$ \citep{serkowski73}. 
The Galactic reddening in the direction of SN 2011hs is E(B-V)$=0.011 \pm 0.002$ \citep{schlafly11}, yielding  $p_{\text{ISP}} < 0.099 \pm 0.018$ percent. 
The polarization of HD215544 is at the upper end of this limit (within uncertainties). 
This is much lower than the observed polarization and the Galactic ISP is therefore a small contribution to the total observed ISP. 

\subsection{Estimating the ISP}
In order to estimate the ISP from our observations, some assumptions must be made.
Any such estimate is inevitably model dependent.
In the following section we present three different methods we employed to determine the ISP. 

\subsubsection{ISP determined from the polarization associated with emission lines}
\label{sec:ispep7_em}
\begin{enumerate}
\item[(i)] Assuming that the supernova polarization is null in the wavelength ranges corresponding to strong emission lines (e.g \citealt{stevance16}) can help us quantify the ISP. 
If we consider the Stokes parameter $q$ (similar equations are used for $u$), the total polarization $q_{\text{tot}}$ across a line will follow:
\begin{equation}\label{equ:qtotFtot}
q_{\text{tot}} F_{\text{tot}} = q_{\text{line}} F_{\text{line}} + q_{\text{cont}} F_{\text{cont}}
\end{equation}

Here $q_{\text{line}}$ is the polarization associated with the emission component, $q_{\text{cont}}$ is the continuum polarization, and the total flux across the line is the sum of the line flux and continuum flux: $F_{\text{tot}} =  F_{\text{line}} + F_{\text{cont}}$. 
Since we assume no intrinsic polarization from the SN in the wavelength ranges corresponding to strong emission line, $q_{\text{line}}$ is effectively $q_{\text{ISP}}$.

We first attempted to extract the ISP from our earliest epochs ($-$3, +2 and +10 days) of spectropolarimetry since they have better SNR. 
Additionally we expect our ISP estimates from the $\mathrm{H\alpha}$ line to be more robust at early dates when He\,{\sc i} is still weak. 
In order to retrieve $q_{\text{line}}$ (i.e. $q_{\text{ISP}}$) from equation \ref{equ:qtotFtot}, we re-arranged to obtain
\begin{equation}\label{ispline}
\frac{q_{\text{tot}} F_{\text{tot}}}{F_{\rm cont}} = q_{\text{line}}\times \frac{F_{\text{line}}}{F_{\text{cont}}} + q_{\text{cont}},
\end{equation}
which is the expression of a straight line whose gradient is $q_{\text{line}}$ and intercept is $q_{\text{cont}}$. 
We then extracted the values of $q_{\text{tot}}$ and $F_{\text{tot}}$ at each wavelength bin in the range associated with the strong emission component. 
In order to find $F_{\text{cont}}$ we visually selected spectral regions of continuum on either side of the emission line and traced a straight line across the spectral feature, giving us $F_{\text{cont}}$ at each wavelength bin. 
By far this process was the most difficult and the greatest source of inaccuracy. 
We plotted $\frac{q_{\text{tot}} F_{\text{tot}}}{F_{\text{cont}}}  $ against $\frac{F_{\text{line}}}{F_{\text{cont}}}$ for each wavelength bin in the emission line range, and fitted the points using Orthogonal Distance Regression (ODR).

We used this technique with the blue iron complex, $\mathrm{H\alpha}$, and Ca\,{\sc ii}. 
In the case of SN 2011hs, the ISP is not constant but a function of wavelength, which in the range of interest ($4500-9300$ \r{A}) is assumed to be linear. 
Hence fitting the values of $q_{\text{isp}}$ against wavelength with a straight line would in principle allow us to retrieve the ISP-wavelength relationship. 
When applying this method to epochs 1 to 3, however, the fits were found to be poor and inconsistent with each other.
This could indicate that our assumptions were inadequate, or it could be the result of the great uncertainties associated with the approximation of the continuum flux. 
\\
\item[(ii)] Another approach is to consider the polarization in the spectral regions corresponding to strong emission lines at a later time, here epoch 7 (+40 days). 
By then intrinsic polarization and continuum flux had significantly decreased, therefore for strong emission lines $F_{\text{line}} \gg F_{\text{cont}}$. 
In this limit $F_{\text{tot}}/F_{\text{line}} \rightarrow 1$ and $q_{\text{line}}F_{\text{line}} \gg q_{\text{cont}}F_{\text{cont}} $.
Consequently, Eq. \ref{ispline} simplifies to $q_{\text{line}}=q_{\text{tot}}$. 
Once again $q_{\text{line}}$ is equivalent to $q_{\text{ISP}}$. 

In order to derive the ISP-wavelength relationship we isolated the Stokes parameter values in the following wavelength ranges: $4970-5020$ \r{A}, $5825-6000$ \r{A}, $6985-7075$ \r{A}, $7255-7345$ \r{A} and $8600-8730$ \r{A}, see Figure \ref{fig:ep7_em_isp}.  
We performed an ODR fit, yielding: 
\begin{equation}\label{eq:qisp_no}
q_{\text{ISP}} = 2.76 (\pm 0.80) \times 10^{-4} \times \lambda - 2.80 (\pm 0.59),
\end{equation}
\begin{equation}\label{eq:uisp_no}
u_{\text{ISP}} = 4.21 (\pm 0.59) \times 10^{-4} \times \lambda - 4.42 (\pm 0.43),
\end{equation}
where $\lambda$ is in \r{A}, the gradients have units of \r{A}$^{-1}$ and the intercepts are unitless. 

When estimating the ratio of line to continuum flux in the spectral region of the calcium infrared triplet (our strongest line), however, we find that F$_{\text{line}}$ \about 3$\times$F$_{\text{cont}}$. 
Consequently, assuming that the potentially polarised continuum flux is negligible in the region of the emission lines is debatable.
We therefore employ another, independent, method to estimate the ISP.

\begin{figure}
	\includegraphics[width=\columnwidth]{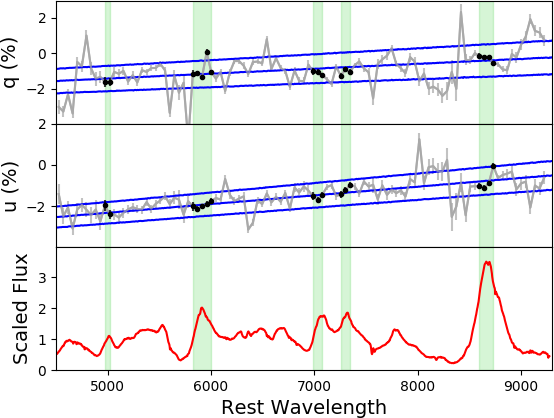}
    \caption{\label{fig:ep7_em_isp}Linear regression (blue) of the Stokes parameters $q$ and $u$ at +40 days in the wavelength range corresponding to emission lines (green highlight). The fitted points are shown in black. The The 1-$\sigma$ lines above and below the fits were determined using Monte Carlo sampling of the gradient and intercept of $q_{\text{ISP}}$ and $u_{\text{ISP}}$ within their errors. The flux spectrum at +40 days is shown in red.}
\end{figure}
\end{enumerate}

\subsubsection{ISP determination from late time data}
\label{sec:ispep7_fit}
\begin{figure}
	\includegraphics[width=\columnwidth]{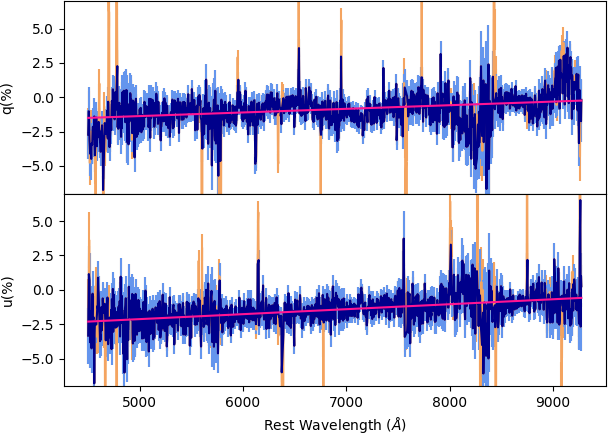}
    \caption{\label{fig:ep7_isp}Stokes parameters $q$ and $u$ at +40 days (dark blue; errors in light blue) 3$\sigma$-clipped according to $\Delta \epsilon$; the discarded points are shown in orange. The bin size for the data is 3.3\r{A}. The fits to the $\sigma$-clipped data are shown in magenta.}
\end{figure}

In the context of electron scattering by the SN ejecta, at late times the polarization intrinsic to the SN will tend towards zero since the electron density decreases as the ejecta expand, and the observed polarization therefore tends toward that of the ISP. 
By assuming complete depolarization at our last epoch (+40 days) and fitting a straight line to the normalised Stokes parameters $q$ and $u$, the wavelength dependent $q_{\text{ISP}}$ and $u_{\text{ISP}}$ can then be found (see Figure \ref{fig:ep7_isp}). 
The best results were obtained when using the original data at +40 days (not rebinned to 45\r{A}), where spurious points were removed according to their corresponding $\Delta \epsilon$.
Data points whose $\Delta \epsilon$ showed a 3$\sigma$ or greater deviation from zero were not included in our fits.

The resulting ISP-wavelength relationships are: 
\begin{equation}\label{eq:qisp}
q_{\text{ISP}}\, {\rm (percent)} = 2.63 (\pm 0.19) \times 10^{-4} \times \lambda - 2.68 (\pm 0.14),
\end{equation}

\begin{equation}\label{eq:uisp}
u_{\text{ISP}} \, {\rm (percent)} = 3.61 (\pm 0.17) \times 10^{-4} \times \lambda - 3.94 (\pm 0.12),
\end{equation}
where $\lambda$ is in \r{A}, the gradients have units of \r{A}$^{-1}$ and the intercepts are unit-less. 
The reduced chi-squared ($\chi^2_{\nu}$) values on the $q$ and $u$ fits were 1.39 and 0.93, respectively. 
The fits are therefore satisfactory. 
Additionally, the parameters derived here are consistent with Eqs. \ref{eq:qisp_no} and \ref{eq:uisp_no}. 
Consequently, we are confident in our ISP estimate.
We chose to use Eqs. \ref{eq:qisp} and \ref{eq:uisp} for ISP removal to avoid the propagation of the large uncertainties associated with Eqs. \ref{eq:qisp_no} and \ref{eq:uisp_no} onto our ISP removed data.
The intrinsic polarization of SN 2011hs is presented and analysed in Section \ref{sec:pol}. 

\subsection{Host galaxy of SN 2011hs and Serkowski law}
\label{sec:serk}
From the upper limits on the Galactic ISP derived in Section \ref{sec:gal_isp} we can conclude that most of the ISP in the data of SN 2011hs originated from IC 5267.
Under the assumption that the size and composition of interstellar dust grains are similar in the host galaxy of SN 2011hs as they are in the Milky Way, the Serkowski limit $p_{\text{ISP}} < 9 \times E(B-V) $ can be applied to derive a value for the maximum $p_{\text{ISP}}$. 
In order to do so, a value for the reddening of IC 5267 must be calculated. 
To this end, we measured and averaged the equivalent widths of the Na\,{\sc i} D line in the data of epochs 1 to 3 (at later epochs the blend of sodium with He\,{\sc i} made our equivalent width measurements unreliable). 
We then used the relationship between equivalent width and reddening described by \cite{poznanski12}, and calculated $E(B-V)_{\text{host}} = 0.11\,(\pm 0.02)$. 
According to \cite{serkowski73}, this yields $p_{\text{ISP}} < 1.09 \,(\pm 0.18)$. 
At 5000 \r{A}, however, we find $p_{\text{ISP}} = 2.50 (\pm0.15)$ percent, as calculated from $q_{\text{ISP}}$ and $u_{\text{ISP}}$ derived in Section \ref{sec:ispep7_fit}. 
Therefore, the assumption that the dust of IC 5267 or in the vicinity of SN 2011hs is similar to that of the Milky Way may not be correct.
It is however worth noting that the Na\,{\sc i}D lines are caused by discrete gas clouds and do not probe the more diffuse ISM, therefore only providing a lower limit on the dust extinction. Consequently, the \cite{poznanski12} estimate may not be completely adequate. 

Another approach is to fit our data at +40 days with the Serkowski law \citep{serkowski75}:
\begin{equation}\label{eq:serk}
p(\lambda) = p_{\text{max}}\exp\big[-K\text{ln}^2\big(\frac{\lambda_{\text{max}}}{\lambda}\big)\big],
\end{equation}
where $K$ is a constant, $p_{\text{max}}$ is the maximum polarization and $\lambda_{\text{max}}$ is the wavelength at polarization maximum. 
We attempted to find new values of the parameters $p_{\text{max}}$ and $\lambda_{\text{max}}$ by minimising the $\chi^2$ to better fit our data.
Two forms of the constant $K$ were considered: the original value derived by \cite{serkowski75}  $K_S=1.15$, and the $\lambda_{\text{max}}$ dependent form of  \cite{whittet92}  $K_W=0.01 + 1.66\lambda_{\text{max}}$.

The normalised Stokes parameters $q$ and $u$ can be expressed as functions of the wavelength dependent degree of polarization $p(\lambda)$ and the polarization angle $\theta$ as
\begin{equation}\label{eq:qu}
q(\lambda) = p(\lambda)\cos(2\theta)\; \text{and} \; u(\lambda) = p(\lambda)\sin(2\theta),
\end{equation}
where $p(\lambda)$ is defined by Eq. \ref{eq:serk} and here $\theta = 120$\degree\, (as from Eqs. \ref{eq:qisp} and \ref{eq:uisp} the polarization angle $\theta$ is found to be 120\degree$\pm2$\degree\, across our wavelength range). 
Note that we did not attempt to remove the Galactic component, since it is $<0.1$ percent and therefore contributes very little to the total ISP.

\begin{figure*}
	\includegraphics[width=17cm]{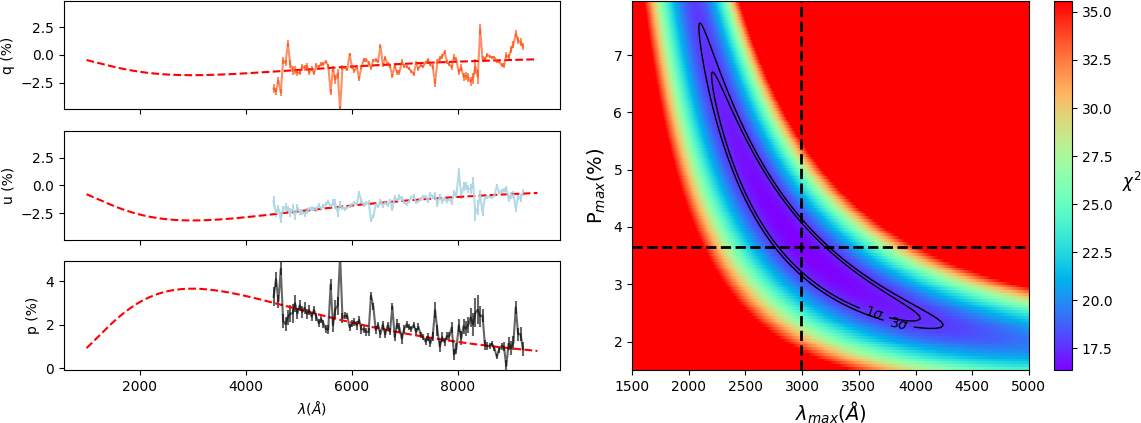}
    \caption{\label{fig:fit_serk}\textbf{Left Panel:} The red dashed lines represent the Serkowski law fits ($K=1.15$ ) to the normalized Stokes parameters q and u at +40 days, and the corresponding fit to $p$. \textbf{Right Panel:} Reduced $\chi^2$ in parameter space. The 1$\sigma$  and 3$\sigma$ contours are plotted as thick black lines, the dashed lines indicate the values of p$_{\text{max}}$ and $\lambda_{\text{max}}$ for which $\chi^2_{\nu}$ is minimized. The colorscale shows the evolution of $\chi^2_{\nu}$ in parameter space. }
\end{figure*}

\begin{figure*}
	\includegraphics[width=17cm]{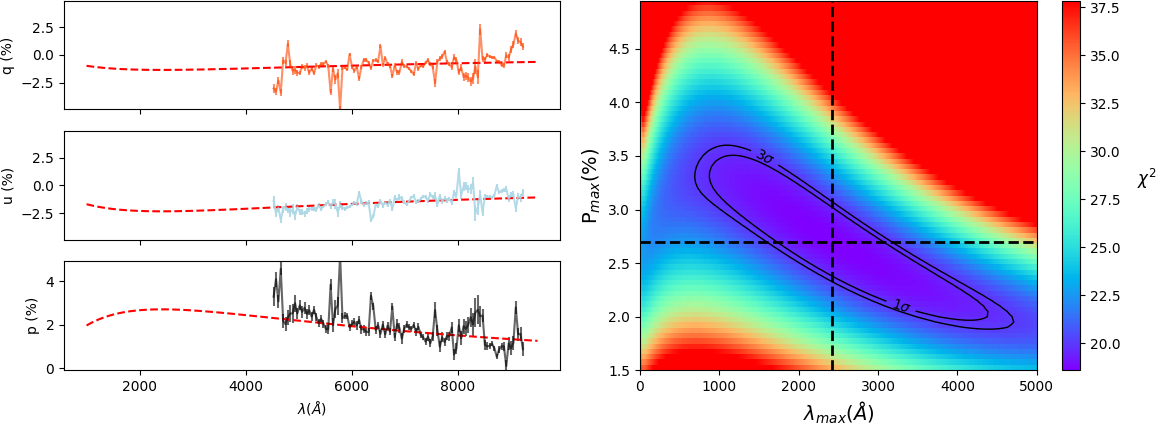}
    \caption{\label{fig:fit_whittet}\textbf{Left Panel:} The red dashed lines represent  Serkowski law fits ($K=0.01 + 1.66\lambda_{\text{max}}$) to the normalized Stokes parameters q and u at +40 days, and the corresponding fit to $p$. \textbf{Right Panel:} Reduced $\chi^2$ in parameter space. The 1$\sigma$  and 3$\sigma$ contours are plotted as thick black lines, the dashed lines indicate the values of p$_{\text{max}}$ and $\lambda_{\text{max}}$ for which $\chi^2_{\nu}$ is minimized. The colorscale shows the evolution of $\chi^2_{\nu}$ in parameter space.  }
\end{figure*}

The fits to $q$, $u$ and $p$ performed using $K_S$, as well as  values of $\chi^2_{\nu}$ in parameter space are shown in Figure \ref{fig:fit_serk}. 
The values of the reduced chi-squared (16.4 at best) indicate a relatively poor fit, and the size of the 1 and 3$\sigma$ contours in Figure \ref{fig:fit_serk} reflect the uncertainty on the best parameter values. 
Nevertheless, it is possible using the 3$\sigma$ contours to put limits on p$_{\text{max}}$ and $\lambda_{\text{max}}$.
We find that 2090\r{A} $<\lambda_{\text{max}}<$ 4245\r{A} and 2.24 percent $<$ p$_{\text{max}}<$ 7.55 percent, with 99.7 percent confidence. 

We also fitted a Serkowski law with $K_{W}$ \citep{whittet92}, as shown in Figure \ref{fig:fit_whittet}, along with the plot of $\chi^2_{\nu}$ in parameter space. 
Once again, the reduced chi-squared shows that the fit is relatively poor and the 1$\sigma$ contour covers a large portion of parameter space. 
We can however put limits on the values of p$_{\text{max}}$ and $\lambda_{\text{max}}$ from the 3$\sigma$ contour: 690\r{A} $<\lambda_{\text{max}}<$ 4700\r{A} and 1.88 percent $<$ p$_{\text{max}}<$ 3.66 percent with 99.7 per cent confidence. 

\subsection{Applicability of Serkowski's law}
\label{sec:disc_serk}

The $\chi^2_{\nu}$ values across parameter space showed very large 1$-\sigma$ contours, encompassing many p$_{\text{max}} - \lambda_{\text{max}}$ pairs.
This is mostly due to the fact that the  $\lambda_{\text{max}}$ peak is not in the  wavelength range covered by our observations, hence making it difficult to better constrain the fit. 
The limits placed on the values of $\lambda_{\text{max}}$ extended well beyond the optical range into the far UV.
This is a concern as Serkowski's law was empirically defined from optical data \citep{serkowski75}, and it has been found that the law is not necessarily applicable in the UV (e.g. \citealt{anderson96, martin99, patat15}).
Furthermore \cite{martin99} explain that the Serkowski equation (see Eq. \ref{eq:serk}) is not flexible enough to provide adequate fits all the way from UV to IR for a given $K$, and that often performing a three parameter fit to find p$_{\text{max}}$, $\lambda_{\text{max}}$ and $K$ gives better results. \cite{patat15} applied this method, and for the case of SN 2006X found $K=1.47\pm0.05$, which is very different from $K_{\text{W}}$ and $K_{\text{S}}$ which we used for our fits.
Therefore, fitting data which peak in the UV with optical Serkowski laws, as is done here, may not be adequate.

The very different shapes of the $\chi^2_{\nu}$ contours in parameter space for the two values of $K$ used in this work (see Figures \ref{fig:fit_serk} and \ref{fig:fit_whittet}) shows the importance of the choice of $K$. 
Given the nature of our data, in particular the fact that $\lambda_{\text{max}}$ is visibly outside of the range we cover, a three parameter fit cannot help constrain our estimate of the parameters.

\begin{figure}
	\includegraphics[width=8cm]{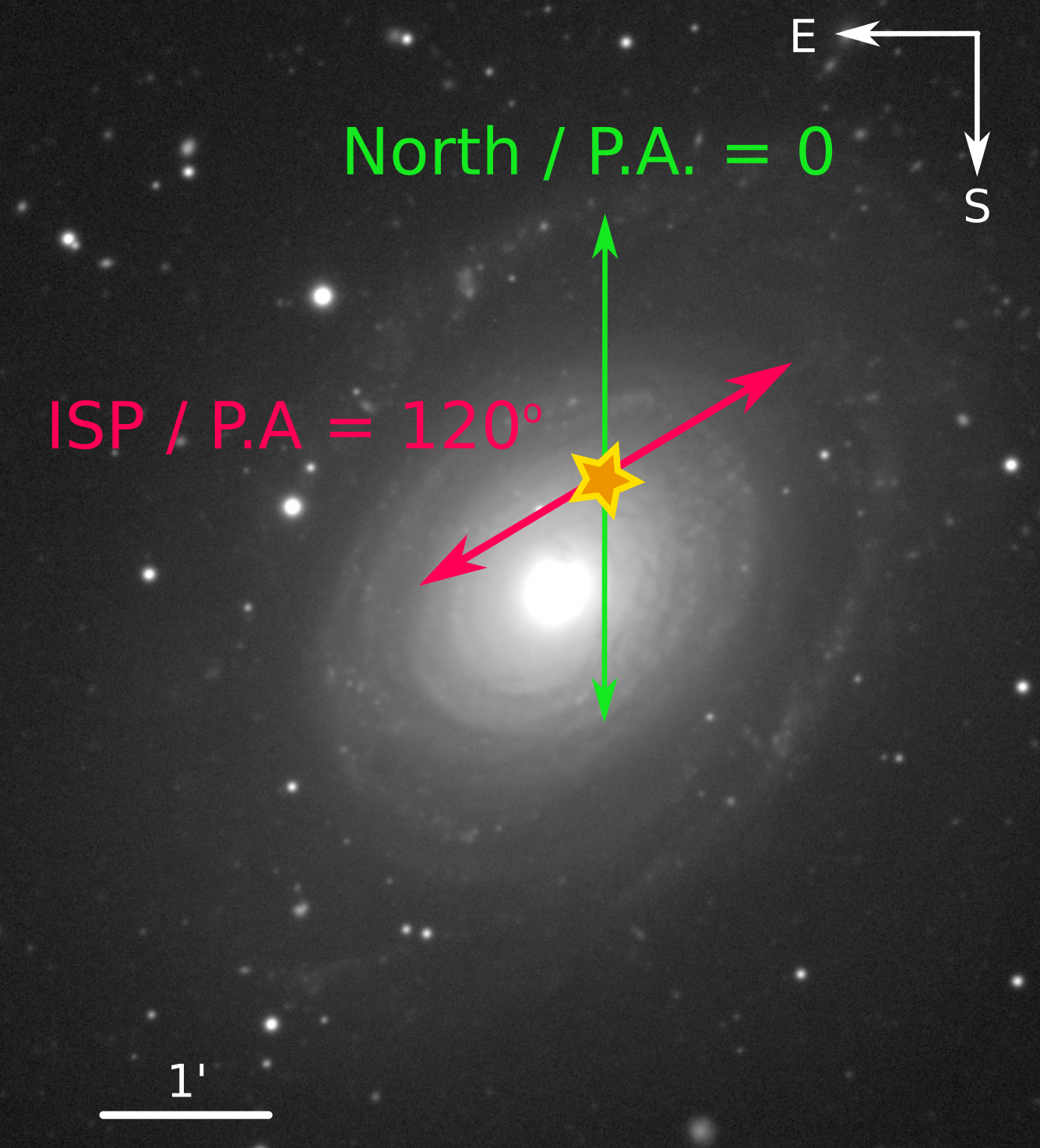}
    \caption{\label{fig:ic5267} Orientation of the ISP polarization angle at the location of SN 2011hs (marked by the star) in IC 5267. The image of the host galaxy is a B, V, R, I composite from the Carnegie-Irvine Galaxy Survey (CGS -- \citealt{ho11}).}
\end{figure}

On the whole, the observed polarization curves are very different from what is expected from Galactic-type ISP. 
Furthermore, it is anticipated that the ISP P.A. should be parallel to the local spiral arm of the host galaxy at the location of the SN, due to the alignment of the dust with the magnetic field of the host. 
However, the ISP angle of \about 120\degree\, is found not to be parallel to the local spiral arm (see Figure \ref{fig:ic5267}).
This could suggest that the dust in IC 5267 or in the local neighbourhood of SN 2011hs maybe different from that in the Milky Way.

An alternative explanation to the observed ISP profile and P.A. is the presence of circumstellar dust. 
Polarization curves increasing at blue wavelengths and with $\lambda_{\text{max}} < 4000$ \r{A} have been see in Type Ia SNe (e.g 1986G, SN 2006X, SN 2014J -- \citealt{patat15}). 
\cite{hoang17} modelled the observed polarization curves and extinction and found this behaviour to be the result of an enhancement in small silicate grains in the dust. 
They suggest this could be due to cloud-cloud collisions resulting from SN radiation pressure.
Alternatively \cite{hoang18} show that large silicate grains can be destroyed by strong radiation fields such as those around massive stars and SNe.
Additionally, \cite{cikota17}, remarked that the polarization curves of Type Ia SNe are similar to those of proto-planetary nebulae, that are produced by scattering in the circumstellar medium. 
Consequently, the peculiar ISP found in the observations of SN 2011hs may hint at the presence of circumstellar dust around the SN.

\section{Intrinsic polarization}
\label{sec:pol}
\begin{figure*}
	\includegraphics[width=17cm]{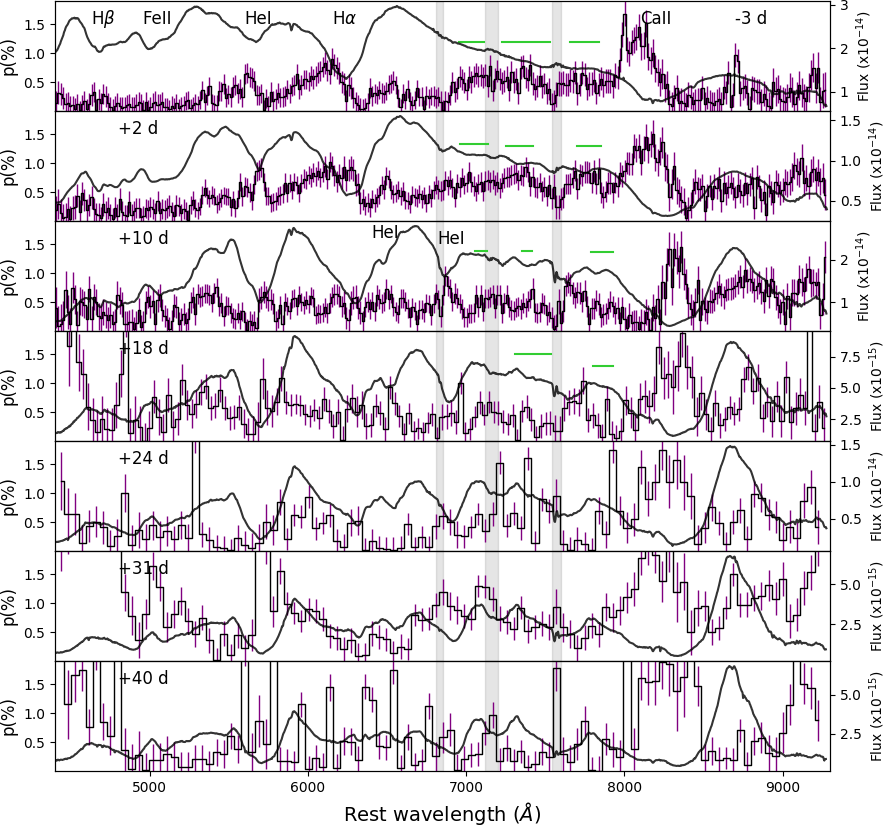}
    \caption{\label{fig:pol} Degree of polarization (corrected for ISP) of SN 2011hs  ranging from $-$3 days to +40 days (purple). The flux spectrum is shown in black, the light grey shaded areas highlight the regions of telluric lines, and the green horizontal lines show the wavelength ranges used to calculate the continuum polarization. The data at $-$3, +2 and +10 days were binned to 15\r{A}, the data at +18 days were binned to 30\r{A} and the following epochs were binned to 45\r{A}. Note that the strong peaks of polarization near 4600\r{A} and 5700\r{A} at +31 and +40 days are not considered to be real (see Section \ref{sec:obs_pol}).}
\end{figure*}

The degree of polarization of SN 2011hs after ISP correction (see Section \ref{sec:isp}) is shown alongside the flux spectrum in Figure \ref{fig:pol}.

\begin{table*}
\caption{\label{tab:pol_table} Summary of the polarization and polarization angle (P.A.) of the continuum and strong lines of SN 2011hs by epoch.}
\begin{tabular}{ r  c l  }
\hline 
Epoch &  p in percent  (wavelength range) & P.A. \\

\hline
\multicolumn{3}{c}{Continuum} \\

$-$3 days  & 0.55 $\pm$ 0.12 (6955-7115 / 7225-7530 / 7655-7840) & 170\degree$\pm$7\degree \\  
+2 days  & 0.75 $\pm$ 0.11 (6960-7140 / 7250-7425 / 7700 -7850) & 158\degree$\pm$4\degree\\  
+10 days & 0.48 $\pm$ 0.09 (7055-7130 / 7350-7415 / 7790-7930) & 141\degree$\pm$7\degree  \\  
+18 days & 0.29 $\pm$ 0.18 (7310-7535 / 7800-7925) & 120\degree$\pm$9\degree \\ 

\multicolumn{3}{c}{He\,{\sc i}$\lambda 5876$} \\

$-$3 days& 0.59 $\pm$ 0.05 (5590-5670) & 173\degree$\pm$8\degree \\  
+2 days& 0.89 $\pm$ 0.03 (5660-5715) & 155\degree$\pm$4\degree \\  
+10 days&  0.57$\pm$ 0.05 (5685-5771) & 141\degree$\pm$10\degree  \\  
+18 days&  1.08 $\pm$ 0.36 (5712) & 54\degree$\pm$3\degree \\ 

\multicolumn{3}{c}{$\mathrm{H\alpha}$} \\

$-$3 days&  0.75 $\pm$ 0.08 (6070-6200) & 5\degree$\pm$6\degree \\  
+2 days& 0.89 $\pm$ 0.10 (6085-6160) / 0.87 $\pm$ 0.08 (6225-6295) & 167\degree$\pm$1\degree / 168\degree$\pm$3.4\degree \\  
+10 days& 0.62 $\pm$ 0.02 (6150-6200) / 0.59 $\pm$ 0.01 (6285-6305)  & 155\degree$\pm$3\degree / 163\degree$\pm$1\degree \\  

\multicolumn{3}{c}{Ca\,{\sc ii}} \\

$-$3 days& 1.17$\pm$ 0.20 (8030-8180) & 177$^{\circ}\pm$4$^{\circ}$ \\  
+2 days& 1.25$\pm$ 0.15 (8055-8250) & 166$^{\circ}\pm$4$^{\circ}$ \\  
+10 days& 1.20$\pm$ 0.25 (8275-8375) & 134$^{\circ}\pm$7$^{\circ}$ \\  
+18 days& 1.35$\pm$ 0.30 (8205-8415) & 134$^{\circ}\pm$13$^{\circ}$ \\ 
+24 days& 1.30$\pm$ 0.26 (8100-8400) &  75$^{\circ}\pm$23$^{\circ}$ \\ 
+31 days& 1.79$\pm$ 0.22 (8100-8300) & 41$^{\circ}\pm$18$^{\circ}$ \\ 
+40 days& 1.64$\pm$ 0.16 (8100-8300) &  107$^{\circ}\pm$19$^{\circ}$ \\ 
\hline
\end{tabular}
\end{table*}

We derived the degree of polarization and P.A. for the prominent polarization features and the continuum where possible. 
Each value in this table is associated with a particular wavelength or wavelength range. 
In the former case, we simply recorded the polarization at the corresponding wavelength bin, whereas when a range is provided we averaged the values within that range and used the standard deviation as the error. 
In order to estimate the continuum polarization, we average the polarization values in spectral regions devoid of strong lines and telluric lines.
We could only find suitable regions of the spectra in epochs 1 to 4, and the wavelength ranges considered are indicated by the green lines in Figure \ref{fig:pol}.
The precise ranges used, and the continuum polarization and P.A. derived can be found in Table \ref{tab:pol_table}.

The continuum polarization is found to be constant within errors at epochs 1 and 2, with $p = 0.55\pm0.12$ percent and $p = 0.75\pm0.11$ percent, respectively.
At those dates the P.A. of the continuum is also constant within the error (170\degree $\pm$ 7\degree and 158\degree $\pm$ 4\degree at $-$2 and +3 days, respectively).
The degree of polarization then decreases to $p = 0.48\pm0.09$ percent by +10 days and $p = 0.29\pm18$ percent by +18 days, and a significant rotation of the P.A. is also observed (141\degree $\pm$ 7\degree and 120\degree $\pm$ 9\degree at +10 and +18 days, respectively).

Prominent polarization features of He\,$\lambda5876$ are seen in Figure \ref{fig:pol} from $-$3 days to +18 days with degree of polarization as high as $p=1.08\pm0.36$ percent at the latter epoch.
The P.A. of helium follows that of the continuum at the first 3 epochs (see Table \ref{tab:pol_table}), but a sudden \about90\degree rotation is seen from +10 days to +18 days (from 141\degree to 54\degree, respectively).

The $\mathrm{H\alpha}$ polarization at $-3$ days and +2 days exhibits a very broad feature  extending from \about 5700\r{A} to \about 6300\r{A}.
The broad hydrogen profile rises to $p=0.75\pm0.08$ percent  in the first epoch (see Figure \ref{fig:pol} and Table \ref{tab:pol_table}).
By +2 days  two distinct peaks can be seen centred around 6125\r{A} and 6260\r{A}, which share the same degree of polarization within errors ($p=0.89\pm0.10$ percent and  $p=0.87\pm0.08$ percent, respectively). 
These two peaks become better defined at +10 days, as the underlying broad profile fades, although their amplitude is seen to decrease by \about 0.3 percent with respect to epoch 2. 
In subsequent epochs, no significant hydrogen feature is detected above the noise. 
Over time  $\mathrm{H\alpha}$ exhibits a decrease in P.A., as seen with helium and the continuum.

Calcium shows by far the most prominent polarization features at all epochs, and is the only element to continue exhibiting polarization above noise levels up to our last epoch at +40 days (see Figure \ref{fig:pol} and Table \ref{tab:pol_table}). 
The amplitude of the feature remains roughly constant from $-$3 days to +24 days, between 1 and 1.5 percent, although the P.A is seen to slowly decrease overtime.
In the last two epochs, however, the polarization degree is well above the 1.5 percent level, and a change in the evolution of the P.A is also seen, as between +31 days and +40 days it increases by ~60\degree.

\subsection{Polar Plots}

\begin{figure*}
	\includegraphics[width=17cm]{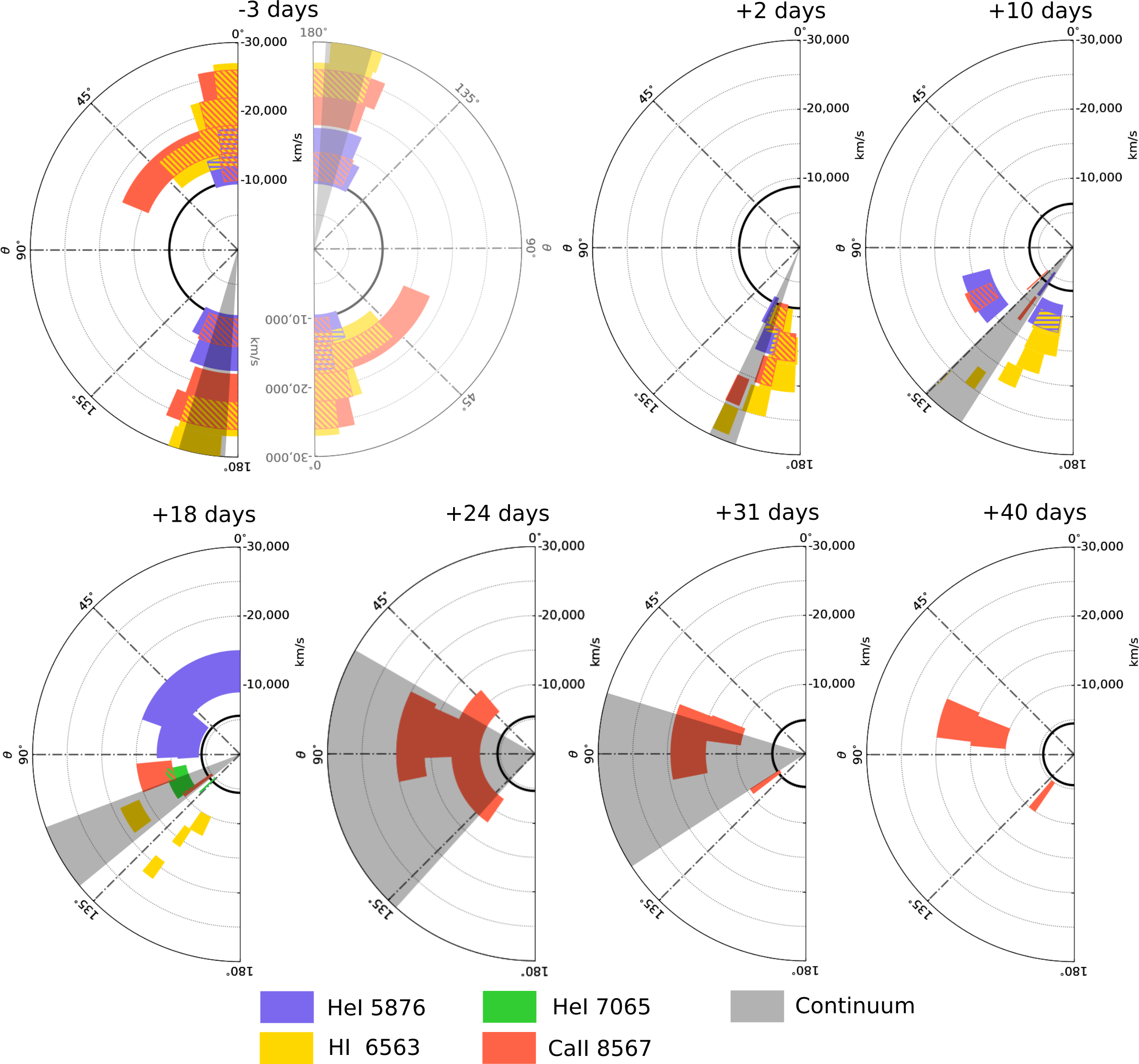}
    \caption{\label{fig:polplot}  Polar plots of SN 2011hs from $-$3 days to +40 days with respect to V-band maximum light. The polarization angles of He\,{\sc i}, $\mathrm{H\alpha}$ and Ca\,{\sc ii} are shown as a function of radial velocity. The velocity of Fe\,{\sc ii}$\lambda 5169$ (proxy for photospheric velocity) is shown in black. The hatched areas represent the overlap between elements. The continuum polarization angle is shown in grey. It should be noted that the apparent symmetry seen at $-$3 days is a consequence of the nature of the Stokes parameters being quasi-vectors. The upper and lower arcs are joined at the 0\degree$-$180\degree\, boundary, as shown by the faded addition of the rotated diagram of the $-$3 days plot. }
\end{figure*}

To help visualise the behaviour of the P.A. in velocity space, we create polar plots, which are diagrams showing the relative positions of the P.A. associated with strong polarization features.
They were first proposed by \cite{maund08D}, and use the radial velocity position and the angle on the plane of the sky as coordinates. 
In Figure \ref{fig:polplot} we show the polar plots at all epochs for helium, hydrogen and calcium.
Note that the data were binned for better visualisation; the centres of the bins are the average P.A. calculated within the range.
The angular extent of the bin on the polar plot represents the error, so that a larger angular extent shows greater uncertainty on the P.A. for a particular line, not a greater physical extent. 

The locations of the continuum shown on Figure \ref{fig:polplot} for epochs 1 through 4 correspond to the polarization angles calculated for the continuum values given in Table \ref{tab:pol_table}.
Unfortunately the continuum polarization could not be formally calculated at the last 3 epochs since we could not isolate spectral regions devoid of strong lines. 
We attempted to estimate the polarization angle of the continuum by identifying its possible locus in $q-u$ space (see Section \ref{sec:loops} and Figure \ref{fig:qu_all}) by using a $2$-$\sigma$ clipping method to eliminate outliers and the contribution of strong lines. 
This method was successfully used for the data at +24 and +31 days, yielding angles of 99\degree$\pm$39\degree and 98\degree$\pm$25\degree, respectively. 
This is consistent with an absence of rotation from +18 days to +24 and +31 days.
At the last epoch the average values of $q$ and $u$ and their standard deviations are consistent with null continuum polarization.

The change in P.A. of the continuum described in the previous section is visualised as a progressive clockwise rotation that takes place at epochs 2 to 4.
Additionally, the P.A. of helium, hydrogen and calcium clearly follow the clockwise rotation of the continuum over time, apart from He\,$\lambda5876$ at +18 days.
In Section \ref{sec:pol} we highlighted the 90\degree rotation of this feature from +10 days to +18 days.
This behaviour is clearly seen in Figure \ref{fig:polplot}, as at +18 days the He\,$\lambda5876$ bins now fall into the upper quadrant, when the helium data were located in the bottom quadrant at +2 and +10 days.
It is also very distinct from the behaviour of the other line features and the continuum.

\subsection{$q-u$ plane}
\label{sec:loops}

\begin{figure*}
	\includegraphics[width=19cm]{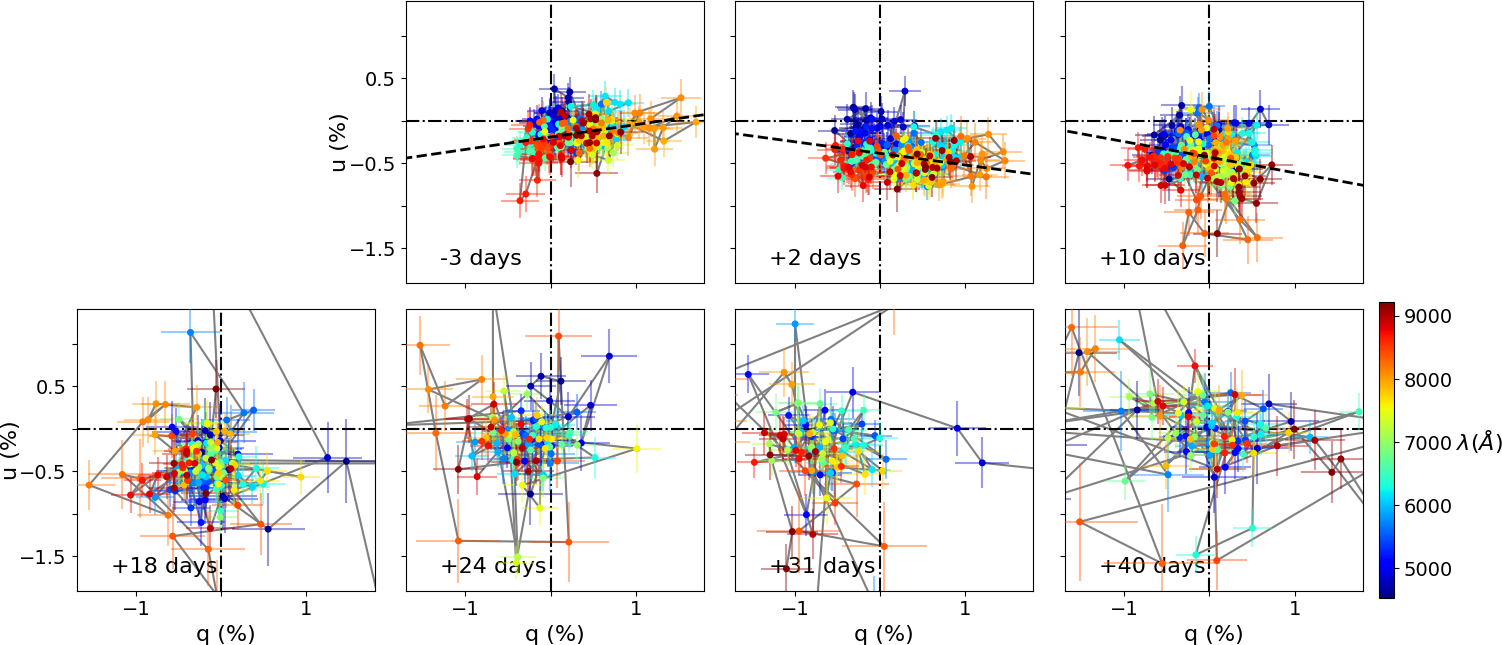}
    \caption{\label{fig:qu_all} Stokes $q-u$ planes of the ISP corrected data of SN 2011hs from $-$3 days to 40 days after V-band maximum. The data were binned to 15 \r{A} and the colour scale represents wavelength. The dominant axes of epoch 1 to 3 are shown as the dashed black line, and were calculated using Orthogonal Distance Regression (ODR). }
\end{figure*}

\begin{figure*}
	\includegraphics[width=15cm]{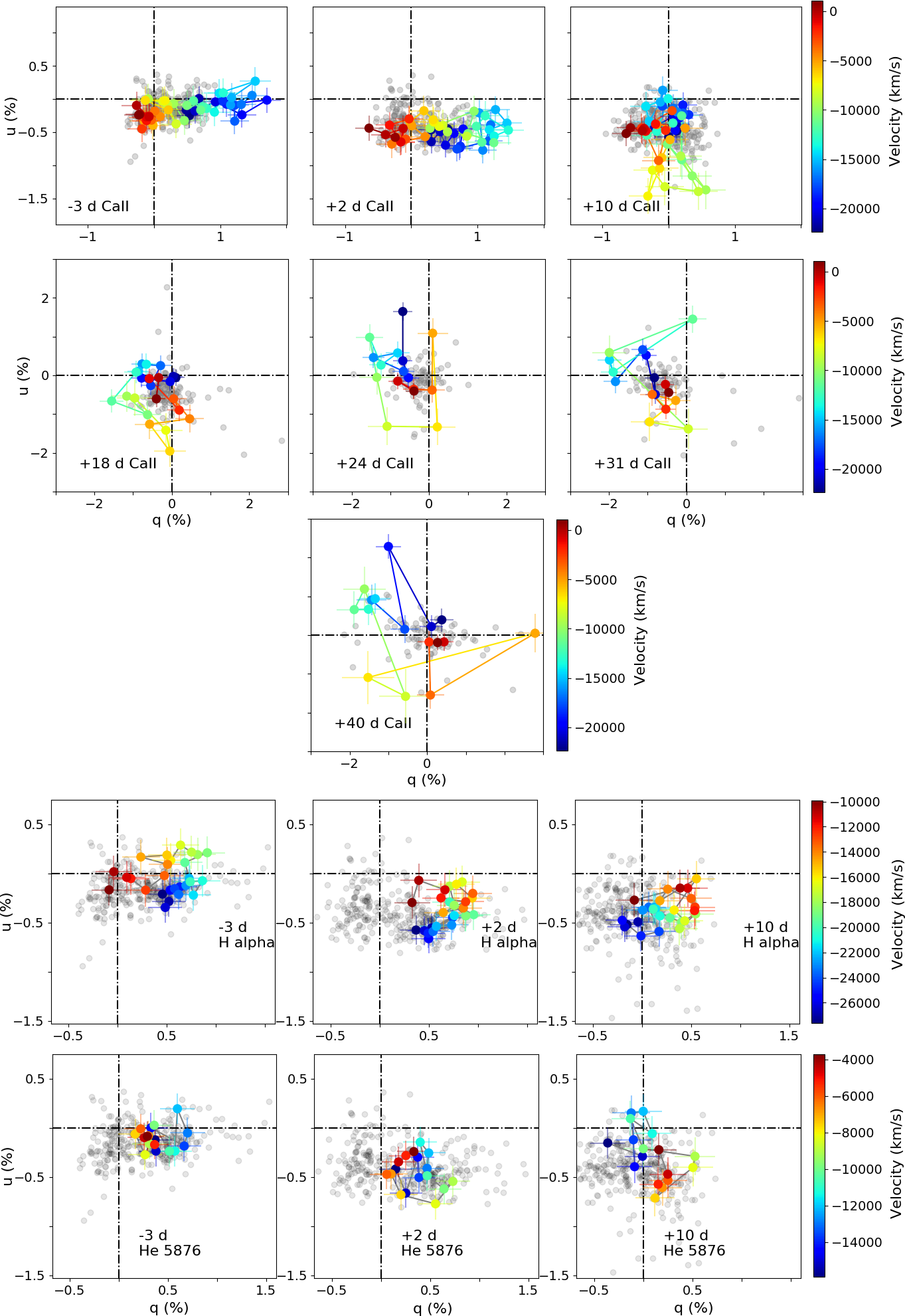}
    \caption{\label{fig:loops} Stokes $q-u$ plots of the ISP corrected data of He\,$\lambda5876$ and $\mathrm{H\alpha}$ and Ca\,{\sc ii}.  The colour scales represent velocity in km\,s$^{-1}$ and vary between species for better visualisation. The grey points show the rest of the data across the whole wavelength range covered by our observations. }
\end{figure*}

In Figures \ref{fig:qu_all} and \ref{fig:loops} we show the ISP corrected polarization data of SN 2011hs plotted on the Stokes $q-u$ plane.
This is a very useful tool to try to understand the geometries of the ejecta \citep{wang01}. 
Bi-axial geometries will result in a linear structure of the polarization data on the $q-u$ plane.
More complex geometries (such as the addition of a third axis of symmetry) will cause departures from such an alignment, sometimes in the form of a smooth rotation of the P.A. across the wavelength range of spectral lines, i.e. loops \citep{WW08}.
Potential physical causes for these geometries and interpretations of the $q-u$ plots will be discussed in Section \ref{sec:pol_origin}.

The spectropolarimetric data of SN 2011hs at $-3$ days, +2 and +10 days show elongated ellipses on the  $q-u$ plane, and the data were fitted with a dominant axis by performing Orthogonal Distance Regression (ODR) over the whole data range. 
The lines of best fit at $-3$, +2 and + 10 days were found to have inclinations on the $q-u$ plane of $8.3\pm1.4$\degree, $172.3 \pm1.4$\degree and $169.6 \pm2.2$\degree, respectively.  
Therefore a clockwise rotation is observed in the orientation of the dominant axis as well as in the continuum P.A. (see Figure \ref{fig:polplot} and Table \ref{tab:pol_table}).

The data at epochs 4 to 7 (+18 to +40 days) have a lower signal-to-noise ratio and show no significant alignment along a dominant axis.
The data at +18, +24 and +31 days present no significant elongation, but their slight offset from the origin of the  $q-u$ plane is due to a residual level of intrinsic polarization. 
At +40 days no elongation is seen and the data are centred  around the origin. which is expected since it was used to quantify the ISP.

In order to look for departures from bi-axial geometry it is worth isolating the spectral regions corresponding to strong lines and seek loops. 
In Figure \ref{fig:loops}, we show the $q-u$ plots of Ca\,{\sc ii}, $\mathrm{H\alpha}$ and He\,$\lambda5876$.
The calcium data at the first two epochs follow a clear linear configuration.
A striking evolution is then seen between +2 days and +10 days, with the appearance of a loop; loops are seen at all subsequent epochs. 
This evolution indicates that at early days the calcium polarization probes a bi-axial geometry and a third axis comes into play at epoch 3, resulting in the loops seen at and after +10 days.
Hydrogen, on the other hand, shows the clearest loop at $-$3 days, where the $\mathrm{H\alpha}$ line is strongest. 
This indicates a break of the bi-axial geometry by hydrogen that is not shared by the calcium.
Lastly, as the helium lines suddenly strengthen in the flux spectrum by +2 days, a loop also becomes visible in the wavelengths associated with He\,$\lambda5876$.
This loop becomes more prominent by +10 days.

The observables presented are discussed and compared to those of other IIb SNe in the following section.

\section{Discussion}
\label{sec:disc}

\subsection{Potential interpretations for the polarization}
\label{sec:pol_origin}

Before we offer potential interpretations for the polarization of SN 2011hs, we summarise how the polarization is understood to probe the ejecta geometry as well as some literature base cases.
We will then place the observational properties described in Section \ref{sec:pol} within this framework.
 
As mentioned in the introduction, the opacity in SN ejecta is dominated by electron scattering, causing the light to be polarised with a vector tangential to the photosphere at the point of scattering. 
After a few hours to a few days with respect to the explosion, the ejecta can be assumed to be in homologous expansion ($r = v \times t$) in all directions.
Consequently, the physical structure of the envelope does not change over time and variations in $p, q, u$ can be understood in terms of a receding photosphere in a hydro-dynamically `inactive' structure.
Additionally, homologous expansion means that different wavelengths across a line feature will probe different ``velocity slices", which are flat projections on the sky.
As a result of the flat projection of a 3D envelope, the outer regions of the projected structure will correspond to regions of the envelope with lower optical depth than the central regions.
Furthermore, the degree of polarization is dependent on optical depth, as demonstrated by the Monte Carlo simulations of \cite{hoflich91}.

In this picture, the integrated polarization of spherical unresolved ejecta is 0, whereas aspherical geometries result in incomplete cancellation of the Stokes parameters and a net polarization signal.
The observed time-scales of variations in $p$ and the polarization angle (P.A.), in combination with the Doppler shift of lines, provide a direct link between the observables and the structure of the envelope.

Polarization probes deviations from sphericity in both the structure of the photosphere and the direction of the radiation reaching the photosphere. 
This leads to three classical base cases:
\begin{enumerate}
\item Aspherical electron distributions, such as an ellipsoidal photosphere, resulting in continuum polarization \citep{1957lssp.book.....V, hoflich91}.
\item Partial obscuration of the underlying Thomson-scattering photosphere leading to line polarization (e.g \citealt{kasen03}).
\item Asymmetric energy input, e.g. heating by off-centre radioactive decay \citep{chugai92, hoflich95}.
\end{enumerate}

Naturally, several or all of these configurations can occur.
These three base cases provide a framework within which to understand changes in the continuum P.A. over time, as well as loops on the $q$-$u$ plane at specific epochs.

Changes in the continuum P.A. over time arise if the geometry of the ejecta is more complex than a simple bi-axial configuration. 
In the case of an ellipsoidal photosphere, for example, the overall evolution of the density gradient over time causes a change from a prolate to an oblate configuration, as the photosphere traverses layers with less steep density structures or a recombination front \citep{hoflich99}. 
This can be seen as a mixture of two case (i) scenarios (oblate + prolate), and this departure from the simple bi-axial geometry will result in a rotation in continuum P.A. across epochs. 
Another possible configuration causing such a rotation is a mixture of case (i) and case (iii), whereby the photosphere has a bi-axial geometry but an off-axis energy source also comes into play, resulting in a tri-axial geometry.

Rotation of the P.A. across spectral lines (loops) at a particular epoch can also be understood as being the result of a tri-axial component. 
In the mixture of cases (i) and (iii) described above (an off-axis energy source in a bi-axial photosphere) the change of P.A. with wavelength arises as a result of the frequency dependence of the thermalisation depth (below which information about the geometry is lost due to multiple scattering -- \citealt{hoflich95}). 
Alternatively, loops will also form if the distribution of the line forming region blocking the photosphere is not aligned with the symmetry axis created by an underlying bi-axial photosphere -- case (i) + case (ii) -- or by an off-centre energy source -- cases (ii) + case (iii).

In Section \ref{sec:pol} the continuum polarization was seen to be significant and constant within errors at the first two epochs ($p = 0.55\pm0.12$ and $p = 0.75\pm0.11$ percent at $-$3 and +2 days, respectively), but then decreases down to \about 0.3 percent by +18 days (see Table \ref{tab:pol_table}).
Simultaneously, the $q-u$ plots show significant elongation in the first 2 epochs, which then becomes less prominent by +10 days finally leading to an absence of a clear dominant axis by +18 days (see Figure \ref{fig:qu_all}).
This suggests that the ejecta initially show significant bi-axial geometry and overall appear more spherical at later dates. 
In the context of an oblate ellipsoid, the early time ejecta differs from sphericity by \about10 percent \citep{hoflich91}.

Now focusing on P.A., we saw in Table \ref{tab:pol_table} and Figure \ref{fig:polplot} that the continuum P.A. showed a gradual clockwise rotation (until +18 days, after which no precise estimate could be made). 
The average P.A. of the polarization features associated with helium, hydrogen and calcium follow the behaviour of the continuum at the first 4 epochs (apart from He\,{\sc i} $\lambda5876$ at +18 days, which we discuss below).
Additionally, a clockwise rotation of the dominant axis is also seen in the $q-u$ planes between epochs 1 and 3.
The homogeneous behaviour of the continuum and line polarization suggests that they are in some way coupled.
In a scenario where line polarization is solely due to partial blocking of the photosphere by a non-isotropic distribution of the line forming regions -- our case (ii) described above -- the observed values in P.A. would indicate that the asymmetries in these line forming regions must individually follow a similar direction to the global asymmetry. 
On the other hand, a scenario in which the line polarisation is the result of global geometry effects requires no serendipitous alignment, and we therefore prefer this interpretation.

In order to see a rotation in polarization angle over time, a tri-axial geometry is required. 
A possible configuration could be an off-axis energy source within initially ellipsoidal ejecta. 
In the framework presented above, this would be a mixture of a case (iii) and a case (i).
As the photosphere recedes through the ejecta, the off-axis energy source would be revealed and add a third axis to the original bi-axial geometry, causing a change in P.A. over time.
Our estimates of the photospheric velocities (see Table \ref{tab:vel_table}) show that at the first two epochs the photosphere is found at $-$9,860 km\,s$^{-1}$ and $-$8,760 km\,s$^{-1}$, respectively.
At epoch 3, when the rotation of the P.A. becomes significant, the photosphere has receded considerably down to $-$6,320 km\,s$^{-1}$, which could have started to reveal a deeper off-centre energy source. 

As previously mentioned, the emergence of an off-axis energy source at +10 days, breaking the initial bi-axial symmetry of the ejecta, would result in the loops on the $q-u$ plane.
Indeed, we see that the calcium feature forms clear lines on the $q-u$ plots at $-$3 and +2 days, and then exhibits a clear loop at +10 days (see Figure \ref{fig:loops}).
This, however, is not seen in hydrogen or helium, which show loops at earlier dates. 
As described in Section \ref{sec:loops}, $\mathrm{H\alpha}$ has the clearest loop at $-$3 days, where hydrogen dominates the spectrum. 
The clarity of this loop diminishes over the next 2 epochs, whereas the He\,{\sc i} $\lambda5876$ loop (non-existent at $-$3 days), becomes more distinct as helium starts dominating the spectrum.
This seems to suggest that in addition to the global geometry effects dominating the P.A. behaviour of the line polarization features, there could be line specific effects.
Therefore, some anisotropies in the distribution of the line forming regions of hydrogen and helium -- case (ii) -- may also be present. 

We want to emphasise that the possible solutions detailed in this section are non-unique. 
Additionally, they do not explain all of the observational characteristics described in Section \ref{sec:pol}. 
The drastic rotation of the P.A. of He\,{\sc i} $\lambda5876$ at +18 days, which differs from the rotation of the continuum and other elements, remains unexplained.
The cause for increase in polarization and rotation of the P.A. by \about60\degree \, in the calcium data at later times is also unclear.
Careful modelling is required, but is beyond the scope of this study.

\subsection{Comparison to previous studies}

\begin{figure}
	\includegraphics[width=8cm]{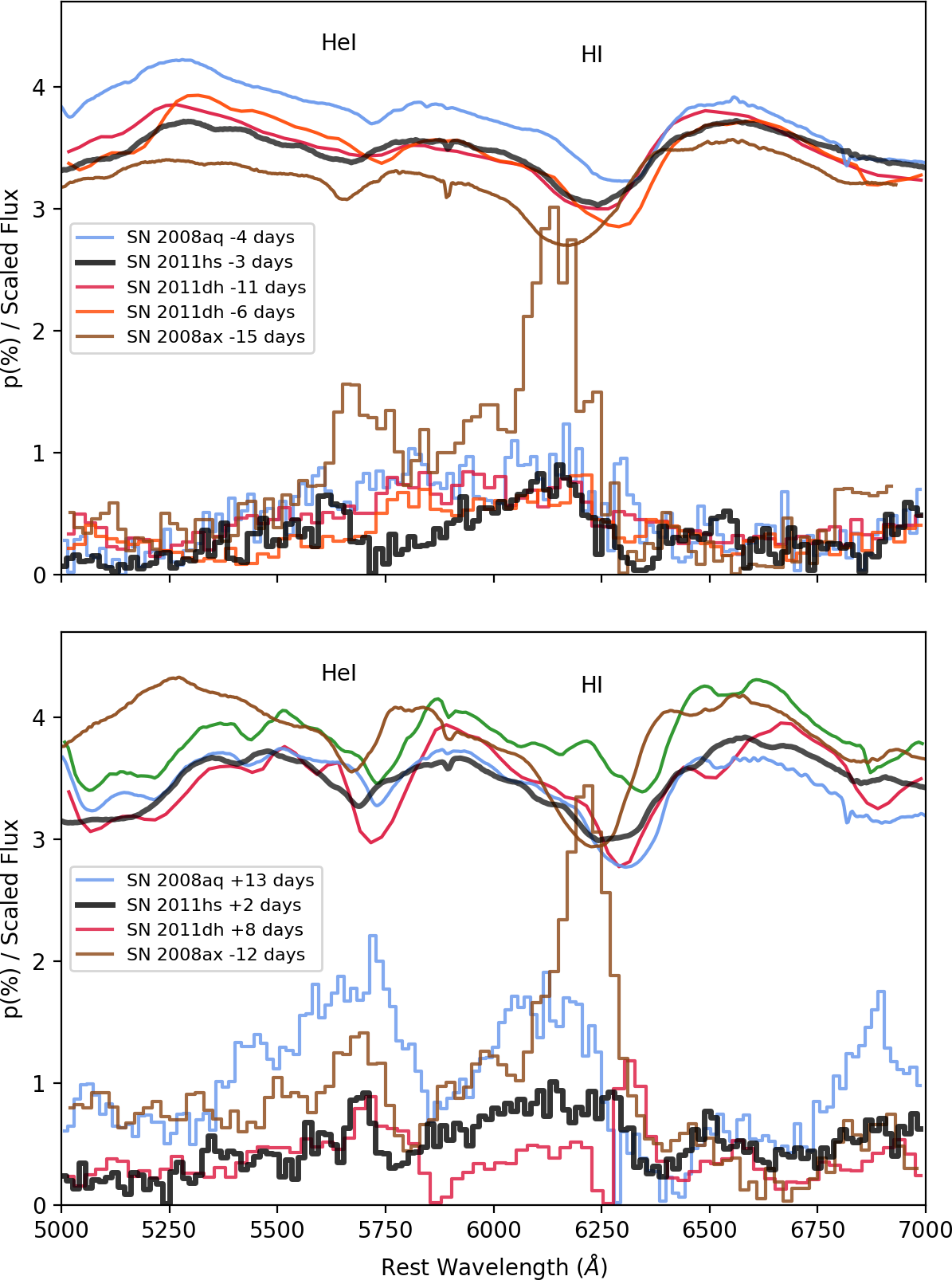}
    \caption{\label{fig:broad_hhe} $\mathrm{H\alpha}$ and He\,{\sc i} $\lambda5876$ polarization features in phases where hydrogen dominates (top panel) and after helium has strengthened (bottom panel). We show SN 2011hs, SN 2008ax \citep{chornock11}, SN 2011dh \citep{mauerhan15} and SN 2008aq \citep{stevance16}. The polarization data are shown as step plots and the flux spectra are plotted for comparison as smooth curves. Note that the phases are quoted with respect to $V$-band maximum instead of explosion date.}
\end{figure}

As previously mentioned, Type IIb SNe are relatively well represented in the spectropolarimetry literature and a variety of spectropolarimetric evolutions have been observed for different objects.
SN 2001ig only showed 0.2 percent continuum polarization at early times, which then rose to \about 1 percent after maximum light \citep{maund01ig}. 
Other cases, exhibited strong continuum polarization early on, with \cite{chornock11} recording 0.64 percent continuum polarization in SN 2008ax 9 days after explosion, or 12 days before $V$-band maximum. 
\cite{stevance16} also found significant continuum polarization (\about 0.7 percent) in SN 2008aq 16 days after explosion, or 4 days before $V$-band maximum. 
Subsequently, the continuum polarization of SN 2008aq rose above 1.2 percent a week after maximum, therefore following a similar pattern to SN 2001ig. 

The increase in polarization as the photosphere probes the deeper ejecta is not universal.
Indeed SN 2011dh \citep{mauerhan15} showed significant continuum polarization (0.45 percent) at 9 and 14 days after explosion (i.e. 11 and 6 days before $V$-band maximum) and decreased to $p<$0.2 percent by 30 days after explosion (i.e a week after $V$-band maximum). 
A similar behaviour is observed in SN 2011hs, where  significant polarization $p = 0.55\pm0.12$ and $p = 0.75\pm0.11$ percent was seen at $-3$ and +2 days with respect to $V$-band maximum, and then decreased by \about 0.25 percent by +10 days, down to \about 0.3 percent at +18 days.

Another common feature of the SN 2011hs and SN 2011dh data is the coupling between the P.A. of the $\mathrm{H\alpha}$ and He\,{\sc i} polarization and that of the continuum.
\cite{mauerhan15} suggested this was best explained by clumpy excitation by plumes of nickel rising from the core.
This could also be the case in SN 2011hs, and would fit in with the scenario of an off-axis energy source described in Section \ref{sec:pol_origin}, where $^{56}$Ni plumes rising from the core could break the global bi-axial geometry of the outer ejecta. 

The presence of an asymmetric nickel distribution could be due to numerous reasons. 
In the case of the jet powered bipolar explosion of SN 1987A, asymmetric nickel distribution was suggested to cause polarization \citep{chugai92}.
Alternatively,  3D simulations by \cite{wongwathanarat13} showed asymmetric distributions of nickel in neutron star kick scenarios, where the iron group elements are ejected in the direction opposite that  of the kick imparted on the remnant. 
These simulations, however, used progenitors of mass 15-20\msol \,at explosion, which is not consistent with the mass limits derived by \cite{bufano14} for SN 2011hs (\mza = 12-15\msol).
This relatively low mass for the progenitor of SN 2011hs could imply the presence of a binary companion. 
Indeed, SN progenitors with \mza $<20$\msol\, require binary interaction to lose enough mass to result in stripped envelope SNe \citep{dessart11}.
It was proposed by \cite{hoflich95} for the case of SN 1993J that an off-centre energy source could be the result of binary interaction, where the asymmetry results from the inner region of the ejecta still being accelerated by the gravitational potential of the binary whereas the faster outer regions are driven far from the orbit of the system very early.
Furthermore, the binary scenario could also explain the early asphericity through binary interaction, similarly to the case of SN 2001ig \citep{maund01ig}.
Note, however, that this is a non-unique solution to our data. 

Certainly one of the most shared characteristics of Type IIb SNe is the presence of line polarization associated with He\,{\sc i} $\lambda5876$ and $\mathrm{H\alpha}$. 
In Figure \ref{fig:broad_hhe} we place SN 2011hs in the context of other SNe at similar spectral stages (where hydrogen dominates and where helium starts becoming prominent). 
We would like to emphasize that the spectropolarimetric behaviour of Type IIb SNe is non-uniform. 
Some SNe (SN 2008aq, SN 2011dh) show a strong blend of the helium and hydrogen polarization features before helium features start strengthening as the photosphere recedes through the ejecta. 
At later dates, once helium lines become stronger, the $\mathrm{H\alpha}$ feature separates from the He\,{\sc i} $\lambda5876$ and two distinct peaks are visible. 
In other cases, e.g. SN 2008ax and SN 2011hs, two distinct peaks of $\mathrm{H\alpha}$ and  He\,{\sc i} $\lambda5876$ can be seen even before helium spectral features start deepening. 
It is interesting that SN 2011hs and SN 2011dh behave differently in this respect, despite the other commonalities in their spectropolarimetric characteristics. 
These disparities should be reproduced by future models of the spectropolarimetry of Type IIb SNe.

The presence of loops associated with the polarization of strong lines has also been observed repeatedly in Type IIb SNe.
One of the most extreme cases was $\mathrm{H\alpha}$ in SN 2008ax (\citealt{chornock11} see their figures 13 and 14). 
In SN 2011hs, we saw loops of $\mathrm{H\alpha}$ and He\,{\sc i} $\lambda5876$ at early times. 
Hydrogen exhibited the most prominent loop at $-3$ days when hydrogen was strongest in the spectrum, whereas the helium loop only arose at +2 days and grew at +10 days as the helium lines strengthened. 
In SN 2011dh the best defined hydrogen loops are also found before helium starts showing prominent features in the flux spectrum (see figure 4 of \citealt{mauerhan15}). 
Contrary to SN 2011hs, however, SN 2011dh shows significant loops in He\,{\sc i} $\lambda5876$ even at the earliest times.
Another major difference between these two objects is in their calcium loops. 
As we saw in Section \ref{sec:loops}, the calcium data in SN 2011hs has a linear structure on the $q-u$ plane around maximum light, and only starts exhibiting a loop 10 days after $V$-band maximum. 
In contrast, SN 2011dh shows its strongest calcium loop at the earliest epoch (9 days after explosion or 11 days before $V$-band maximum).

These differences in the loop behaviour of Type IIb SNe, and the precise origin of the loops of $\mathrm{H\alpha}$ and He\,{\sc i}$\lambda5876$ in SN 2011hs at the first two epochs, are difficult to understand without modelling. 
Toy models have been used in the past (e.g. \citealt{maund05hk}, \citealt{reilly16}), to try to reproduce line polarization and constrain the ejecta geometry.
In an attempt to understand the early hydrogen and helium loops (and potentially extend this to other SNe), we created a model based on the same assumptions, but used a more methodical approach to explore parameter space (Stevance in prep.). 
The result of this work, however, was to demonstrate the great number of degeneracies that arise in such models even when considering a small number of free parameters, as well as the fact that it can
result in good fits to the data even in cases were the original assumptions are invalid. 
This highlights the fact that simplified models must be considered very carefully as they may yield misleading results.

Full hydrodynamic simulations with radiative transfer that can simultaneously reproduce the flux spectrum and the corresponding polarization features will be necessary to better understand the varied geometry of Type IIb and other core-collapse SNe. 
Additionally, higher cadence observations, especially around and soon after maximum light, could help better understand the variations of the ejecta geometry with depth.


\section{Conclusions}
\label{sec:conclusions}
We presented seven epochs of spectropolarimetry for the Type IIb SN 2011hs from $-$3 days to +40 days with respect to $V$-band maximum. 
The observed polarimetry data showed very high levels of polarization increasing towards blue wavelengths (up to \about 3 percent).
We quantified the ISP and identified that most of the observed polarization was caused by the interstellar component.
Fits of the Serkowski law \citep{serkowski75} allowed us to constrain $\lambda_{\rm max}$ to wavelengths $<$ 4245\r{A} or $<$4700\r{A}, depending on the value of K used, either from \cite{serkowski75} or \cite{whittet92}, respectively.
Such levels of ISP have never been observed in a Type IIb SN before. 
Similar behaviours of the interstellar component, with low values of $\lambda_{\rm max}$, have been seen in some Type Ia SNe \citep{patat15}.
This may suggest enhanced levels of small silicate grains, either resulting from cloud-cloud collisions caused by SN radiation pressure, or due to the destruction of large grains by the radiation field \citep{hoang17,hoang18}. 
Consequently, the behaviour and level of the ISP in the spectropolarimetric data of SN 2011hs seem to indicate the presence of dust in the vicinity of the SN, which could reflect the mass loss history of the progenitor. 

The intrinsic polarization of SN 2011hs was retrieved after removal of the ISP component. 
Significant continuum polarization was observed at the first two epochs, with $p=0.55\pm0.12$ percent and $p=0.75\pm0.11$ percent, respectively, corresponding to \about10 percent departure from spherical geometry, in the context of an oblate spheroid \citep{hoflich91}. 
The continuum polarization then decreased by \about 0.25 percent by +10 days and declined further by +18 days.
A strong correlation was found between the behaviour of the P.A. of hydrogen, helium, calcium and that of the continuum, indicating they share a common geometry.
The progressive rotation of the continuum P.A. after epoch 2 can be interpreted as the presence of an off-centre energy source being revealed.
This is supported by the calcium data on the $q-u$ plane, where a dichotomy exists between the first two epochs at which the data form a line (indicating bi-axial geometry), and the following epochs at which loops are observed (evidence for a departure from bi-axial geometry). 

On the other hand, $\mathrm{H\alpha}$ shows the clearest loop at the first epoch when hydrogen is strongest in the spectrum.
It then diminished as He\,{\sc i}$\lambda5876$ starts to show a loop by +3 days, which strengthens by +10 days, where helium is starting to dominate the spectrum.
These characteristics show that the lines of hydrogen and helium probe tri-axial geometries at early times where calcium does not, and therefore line specific geometries must also be a contributor to their polarization. 
A possibility would be the presence of anisotropies in the distribution of their line forming regions.

Compared to previously studied Type IIb SNe, SN 2011hs is most similar to SN 2011dh \citep{mauerhan15}, where a decrease in continuum polarization over time and a correlation between the P.A. of hydrogen, helium and the continuum was observed. 
\citeauthor{mauerhan15} also favoured an off-centre source of energy to explain their observations, namely in the form of plumes of nickel.
This could also be consistent with SN 2011hs, but is not unique and is probably an insufficient solution to our observations.

Lastly, there are a number of features whose origins remain unknown.
At +18 days the P.A. of He\,{\sc i}$\lambda5876$ undergoes a drastic rotation that is not coupled with that of the continuum or other elements.
Additionally, the calcium data at later dates see an increase in polarization and a rotation in P.A. by +60\degree.

On the whole, SN 2011hs brings to our sample of Type IIb SN spectropolarimetry data a number of features that have been previously seen as well as new disparities that need to be explained. 
It is clear from examples such as SN 1993J, SN 2001ig, SN 2008aq, SN 2008ax, SN 2011dh and SN 2011hs that there is great variety in the observed spectropolarimetric, and therefore geometric, properties of Type IIb SNe \citep{tran97, hoflich95, maund01ig, stevance16, chornock11, mauerhan15}.
More observations, especially with a higher cadence around and after maximum, and detailed hydrodynamic models with radiative transfer of the current and future spectropolarimetric data are required for us to better understand the similarities and the diversity in the geometries of Type IIb SNe.

\section*{Acknowledgements}
The authors would like to thank the staff of the Paranal Observatory for their kind support and for the acquisition of such high quality data on the program 088.D-0761. 
This work has made use of data from the European Space Agency (ESA) mission
{\it Gaia} (\url{https://www.cosmos.esa.int/gaia}), processed by the {\it Gaia}
Data Processing and Analysis Consortium (DPAC,
\url{https://www.cosmos.esa.int/web/gaia/dpac/consortium}). Funding for the DPAC
has been provided by national institutions, in particular the institutions
participating in the {\it Gaia} Multilateral Agreement.
We are grateful to S. Parsons for sharing his expertise on white dwarf flux calibration, S. Couch for bringing an insightful reference to our attention, and L. Grimmett for our numerous discussions concerning statistics.
HFS and JB are supported through a PhD scholarship granted by the University of Sheffield.
JCW is supported by the Samuel T. and Fern Yanagisawa Regents Professorship in Astronomy and by NSF Grant 1813825.
AC is supported by grant IC120009 (MAS) funded by the Chilean Ministry of Economy, Development and Tourism, and
by grant Basal CATA PFB 06/09 from CONICYT.
The research of JRM is supported through a Royal Society University Research Fellowship. 
The following packages were used for the data reduction and analysis: Matplotlib \citep{matplotlib}, Astropy \citep{astropy}, Numpy, Scipy and Pandas \citep{scipy}.




\bibliographystyle{mnras}







\bsp	
\label{lastpage}
\end{document}